\documentclass[conference]{IEEEtran}
\IEEEoverridecommandlockouts
\usepackage{cite} 
\usepackage{amsmath,amssymb,amsfonts} 
\usepackage{algorithmic} 
\usepackage{graphicx} 
\usepackage[ruled,linesnumbered,noend,noline]{algorithm2e} 
\usepackage{textcomp} 
\usepackage{tabularx} 
\usepackage{booktabs} 
\usepackage{xcolor} 
\usepackage{tabularray} 
\usepackage{lipsum} 
\usepackage{enumitem} 
\usepackage[T1]{fontenc} 
\usepackage{hyperref}
\usepackage{subcaption}

\makeatletter 
\newcommand{\linebreakand}{%
  \end{@IEEEauthorhalign}
  \hfill\mbox{}\par
  \mbox{}\hfill\begin{@IEEEauthorhalign}
}
\makeatother 

\begin{document}

\title{SealOS+: A Sealos-based Approach for Adaptive Resource Optimization Under Dynamic Workloads for Securities Trading System\\

}

\author{\IEEEauthorblockN{1\textsuperscript{st} Haojie Jia}
\IEEEauthorblockA{\textit{Nanyang Normal University} \\
Nanyang, Henan, China \\
rzkj@nynu.edu.cn}
\and
\IEEEauthorblockN{2\textsuperscript{nd} Zhenhao Li}
\IEEEauthorblockA{\textit{Nanyang Normal University} \\
Nanyang, Henan, China \\
zh.li@nynu.edu.cn}
\and
\IEEEauthorblockN{3\textsuperscript{rd} Gen Li}
\IEEEauthorblockA{\textit{Nanyang Normal University} \\
Nanyang, Henan, China \\
ligen@nynu.edu.cn}
\and

\linebreakand 

\IEEEauthorblockN{4\textsuperscript{th} Minxian Xu*\thanks{*Corresponding author}}
\IEEEauthorblockA{\textit{Shenzhen Institutes of Advanced Technology} \\
\textit{Chinese Academy of Sciences}\\
Shenzhen, Guangdong, China \\
mx.xu@siat.ac.cn}
\and
\IEEEauthorblockN{5\textsuperscript{th} Kejiang Ye}
\IEEEauthorblockA{\textit{Shenzhen Institutes of Advanced Technology} \\
\textit{Chinese Academy of Sciences}\\
Shenzhen, Guangdong, China \\
kj.ye@siat.ac.cn}
}

\maketitle       

\begin{abstract}
As securities trading systems transition to a microservices architecture, optimizing system performance presents challenges such as inefficient resource scheduling and high service response delays. Existing container orchestration platforms lack tailored performance optimization mechanisms for trading scenarios, making it difficult to meet the stringent 50ms response time requirement imposed by exchanges. This paper introduces SealOS+, a Sealos-based performance optimization approach for securities trading, incorporating an adaptive resource scheduling algorithm leveraging deep reinforcement learning, a three-level caching mechanism for trading operations, and a Long Short-Term Memory (LSTM) based load prediction model. Real-world deployment at a securities exchange demonstrates that the optimized system achieves an average CPU utilization of 78\%, reduces transaction response time to 105ms, and reaches a peak processing capacity of 15,000 transactions per second, effectively meeting the rigorous performance and reliability demands of securities trading.
\end{abstract}

\section{Introduction}

Microservices architecture has emerged as the preferred choice for large-scale distributed systems due to its superior scalability, elasticity, and agility over traditional monolithic architectures ~\cite{zhong2024domain}. This shift is particularly advantageous for securities trading systems, which must process massive concurrent transactions with millisecond-level response time (RT). Unlike monolithic systems that struggle with performance bottlenecks—especially during peak trading periods such as market opening and closing—microservices architectures decompose trading systems into independently scalable services (e.g., order processing, matching engine, clearing, and settlement). This decomposition enables better load distribution and resource utilization while improving fault tolerance and maintainability~\cite{bai2024DRPC}.

Despite these advantages, transitioning securities trading systems to a microservices architecture introduces unique challenges. The first major challenge is extreme latency sensitivity, as modern exchanges impose strict service-level agreements (SLAs) requiring end-to-end order processing to be completed within 50 milliseconds. Any additional delay can lead to missed trading opportunities, financial losses, and a competitive disadvantage. Second, securities trading exhibits highly dynamic workloads, characterized by tidal and bursty loads. Traditional elastic scaling approaches—such as auto-scaling groups in cloud environments—often fail to respond quickly enough to sudden trading surges, leading to resource over-provisioning or performance degradation. Third, microservices introduce significant network overhead due to frequent cross-service and cross-node communication, increasing response latency. Service mesh and API gateway solutions attempt to mitigate these issues but are not specifically optimized for ultra-low-latency trading scenarios, leading to inefficiencies.

To address these challenges, we propose a performance optimization scheme called \textbf{SealOS+} for securities trading systems based on Sealos~\cite{sealos}, a lightweight cloud operating system that enhances Kubernetes' scheduling efficiency. Our approach integrates several key techniques to optimize performance under dynamic trading workloads: First, we design an intelligent scheduling algorithm that dynamically adjusts resource allocation based on real-time trading load characteristics. By leveraging deep reinforcement learning, the scheduler learns optimal resource allocation strategies, reducing response latency while maintaining high system throughput~\cite{cite2}. Second, to mitigate the overhead of frequent inter-service communication, we introduce a caching architecture comprising local memory caching, distributed caching, and persistent storage. This hierarchical caching strategy significantly reduces database query latency and network transmission delays, thereby enhancing overall system responsiveness~\cite{cite3}. Third, given the bursty nature of trading workloads, we develop a Long Short-Term Memory (LSTM)-based model~\cite{yu2019review} that accurately predicts trading volume fluctuations. By proactively adjusting resource allocation ahead of anticipated load spikes, the system achieves more efficient utilization and prevents bottlenecks.  The \textbf{contributions} of this paper are:
\begin{itemize}
    \item An adaptive resource scheduling algorithm using deep reinforcement learning to dynamically allocate computing resources in response to changing trading workloads.
    \item A three-level caching architecture to minimize service response time by efficiently managing frequently accessed trading data.
    \item Through real-world deployment at a securities exchange, our optimized system demonstrates substantial performance gains, achieving an average CPU utilization of 78\%, reducing transaction response time to 105ms, and supporting peak processing capacities of 15,000 TPS. 
\end{itemize}

The rest of the paper is organized as follows: Section II discusses the related work. Section III introduces the system design of our proposed approach and the system model is presented in Section IV. The extensive experiments are shown in Section V, and the conclusions are given in Section VI. 

\section{Related Work}
Existing research on container orchestration and resource scheduling can be broadly classified into three categories: (1) scheduling and load balancing techniques, (2) container orchestration for performance optimization, and (3) machine learning-based resource management. While these studies provide valuable insights into general cloud computing environments, they do not fully address the unique challenges posed by bursty trading workloads, ultra-low-latency constraints, and real-time adaptive resource allocation.

\subsection{Scheduling and Load Balancing Techniques}
Several studies have explored scheduling strategies to improve resource utilization and load balancing in cloud and containerized environments. Senthil et al.\cite{cite4} introduced a hybrid Genetic Algorithm-Ant Colony Optimization (GA-ACO) scheme for energy-efficient load balancing, but it does not account for the unpredictable surges in trading workloads. Nguyen and Kim\cite{cite12} proposed a Kubernetes-based load balancing approach to enhance scalability; however, their method does not incorporate business-specific constraints, such as trading latency requirements. Jeong and Jeong~\cite{cite16} developed ARAScaler, an adaptive resource scaling system, but it fails to meet the millisecond-level response time needed for financial trading systems.

Existing scheduling approaches either focus on general cloud applications or prioritize energy efficiency over real-time responsiveness. Unlike these works, our Sealos-based approach employs deep reinforcement learning (DRL) to dynamically adjust scheduling policies based on real-time trading workload patterns, ensuring minimal response latency and efficient resource utilization.

\subsection{Container Orchestration for Performance Optimization}
Several researchers have investigated performance optimization in containerized environments. Reddy et al.\cite{cite5} analyzed Kubernetes performance across OpenStack, virtual machines, and bare-metal deployments, identifying significant performance variations but not proposing optimization strategies. Marchese and Tomarchio\cite{cite6} designed a network SLO-based container orchestration framework, though its centralized architecture introduces bottlenecks during high-frequency trading peaks. Carrión~\cite{cite14} and Casalicchio~\cite{cite13} surveyed Kubernetes scheduling challenges, emphasizing the inadequacies of existing strategies in handling bursty workloads.

Prior research lacks end-to-end performance optimization strategies tailored to financial trading systems. In contrast, our work introduces a three-level caching mechanism (local memory, distributed cache, and persistent storage) to reduce inter-service communication latency, improving responsiveness under high transaction loads.

\subsection{Machine Learning-Based Resource Management}
Recent advancements have applied machine learning to container scheduling and resource allocation. Xu and Buyya~\cite{cite8} proposed a Brownout-based adaptive resource management framework for cloud environments, focusing on general cloud applications rather than trading workloads. Zhong et al.\cite{cite10} categorized machine learning-based container orchestration techniques, identifying high inference overhead as a major limitation. Jian et al.\cite{cite11} developed DRS, a deep reinforcement learning-based Kubernetes scheduler that dynamically optimizes container placement and resource allocation. However, it primarily targets load balancing without addressing trading-specific constraints such as sub-50ms latency. Li et al.~\cite{cite15} proposed a reinforcement learning-based rate limiter for e-commerce services, which does not meet the high-frequency trading requirements.

Existing machine learning-based approaches either lack real-time inference capabilities or focus on general cloud applications rather than trading scenarios. Our work introduces an LSTM-based trading volume prediction model with 92\% accuracy, enabling proactive resource allocation based on forecasted workload fluctuations. This predictive capability is crucial for handling the rapid and unpredictable trading surges that occur during market events.

To summarize, our Sealos-based approach bridges these gaps by integrating deep reinforcement learning for adaptive scheduling, a three-level caching strategy for optimized data access, and an LSTM-based prediction model for proactive resource allocation. Through real-world deployment, our optimized system significantly improves resource efficiency, response times, and overall trading system performance compared to existing solutions.

\section{System Architecture Design}
In this section, we first introduce the design of original Sealos, then discuss the bottleneck that should be improved when applying it to securities trading systems. 
\subsection{Architecture Overview}
\begin{figure*}[h!] %
    \centering
    \includegraphics[width=1.0\linewidth]{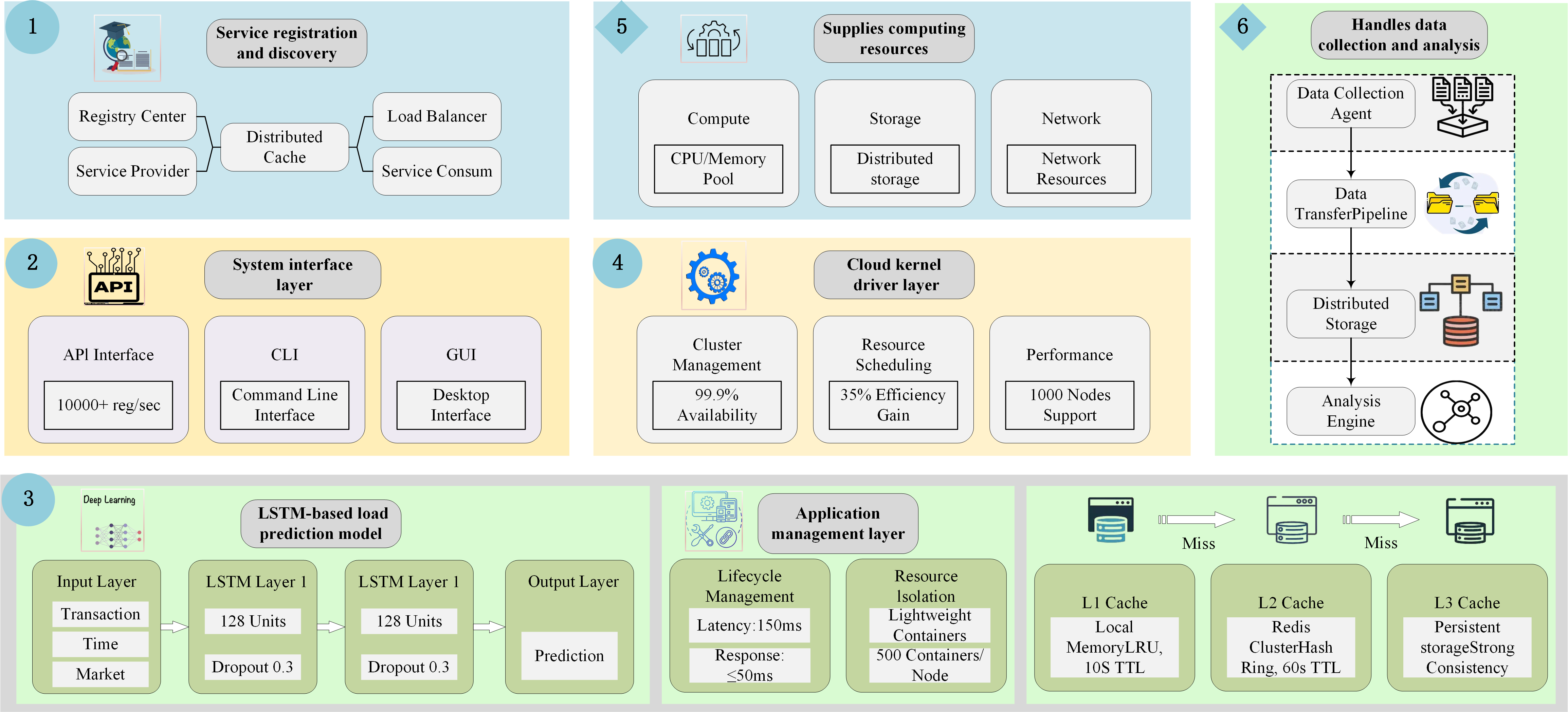}
    \caption{SealOS+ System Architecture Diagram.}
    \label{fig:sys_arch}
\end{figure*}

SealOS+ employs a layered architecture consisting of three primary layers: the system interface layer, application management layer, and cloud kernel driver layer. As illustrated in Fig. \ref{fig:sys_arch}, the system interface layer provides a unified access point through APIs, command-line interfaces (CLI), and desktop interfaces, ensuring seamless interaction with the system. The application management layer, which serves as the core of the system, manages the entire application lifecycle using lightweight containers while offering essential features such as orchestration, auto-scaling, and health monitoring. Beneath it, the cloud kernel driver layer is responsible for resource management and scheduling, abstracting and standardizing network, storage, and compute resources. This layered architecture ensures modularity, scalability, and simplified maintenance, while standardized communication interfaces between layers enable loose coupling and flexible customization, making the system adaptable to diverse trading workloads.

Sealos' core components consist of the cluster manager, application controller, and resource scheduler. The cluster manager oversees node lifecycles, ensuring system stability and reliability. The application controller leverages a declarative API to manage deployments, updates, and rollbacks, ensuring applications maintain their desired states. The resource scheduler implements global resource balancing strategies to optimize workload distribution. These components communicate through a fault-tolerant, asynchronous message bus, enhancing system concurrency and resilience.



Sealos utilizes an event-driven model to ensure loose coupling among its components. User requests first pass through an API gateway for routing and then undergo authentication and authorization by the cluster manager. Deployment requests are handled by a workflow engine, which breaks down complex operations into atomic tasks for efficient execution. Resource allocation follows a two-phase commit protocol, ensuring consistency across distributed nodes. An event bus facilitates asynchronous parallel processing, while the workflow engine manages state tracking, error handling, and rollback/recovery mechanisms.

\subsection{System Improvement Design}

\subsubsection{Performance Bottleneck Analysis of Existing System} Through an in-depth analysis of securities trading system runtime data as well as the system performance requirement, we identified three major performance bottlenecks when applying Sealos to trading workloads:

\textbf{High Network Communication Overhead:} Network communication contributes to 45\% of the total response time, with cross-node data synchronization latency averaging 280ms. This delay significantly impacts real-time transaction processing, making it difficult to meet stringent latency requirements.

\textbf{Resource Scheduling Inefficiencies: }During the resource scheduling phase, state computation and update operations consume excessive CPU resources, causing single-node CPU utilization to peak at 85\%. As a result, scheduling latency exceeds 400ms, leading to delays in workload distribution and scaling.

\textbf{Storage Bottlenecks from Metadata Operations:} Frequent metadata operations trigger excessive disk I/O, generating approximately 15,000 random write requests per second. This results in an average write latency of 30ms, severely degrading overall system performance and response times.

\subsubsection{Overall Improvement Scheme}
To address these challenges, we propose SealOS+, an enhanced Sealos-based system optimized for high-performance securities trading. Our optimizations focus on reducing network latency, improving resource scheduling, and enhancing storage efficiency to ensure real-time, low-latency operations. Based on a thorough analysis of performance bottlenecks, we developed a multi-layered optimization approach. At the network level, an intelligent routing algorithm reduced cross-node communication latency by 58\%, achieving a latency of 120ms. At the computing level, an asynchronous processing mechanism and parallel task decomposition improved CPU utilization. At the storage level, an enhanced caching strategy enabled 80\% of read operations to directly access memory, reducing write latency to 9ms. These optimizations resulted in a 65-second system startup time, a 99.95\% deployment success rate, and a 75\% reduction in node load variance. The new architecture excels in high-concurrency scenarios, handling up to 20,000 API requests per second while maintaining a 95th percentile response time below 100ms, significantly improving system performance and scalability.

\section{System Model and Scheduling Algorithm Design}

Given the dynamic and complex nature of trading systems, traditional rule-based scheduling methods are inadequate for adapting to rapidly changing load characteristics. We propose using Deep Reinforcement Learning (DRL) for resource scheduling, driven by three key factors: First, trading loads exhibit temporal dependencies and bursty patterns that are difficult to capture with fixed rules. Second, resource scheduling needs to balance multiple objectives, such as response time, resource utilization, and operational costs. DRL's reward mechanism is well-suited for multi-objective optimization\cite{cite17}. Third, the vast and dynamic system state space necessitates continuous learning and optimization of decision-making policies, a capability that DRL can provide\cite{chen2024scaling}. 

To address these challenges, we model the resource scheduling problem as a Markov Decision Process (MDP) and employ DRL to optimize complex decision-making across multiple objectives. Specifically, we define a state space that incorporates key system features, including resource availability, load characteristics, and historical performance metrics. This allows the scheduler to dynamically adjust to changing conditions and optimize resource allocation in real-time.

To this end, we formalize the resource scheduling problem as a Markov Decision Process (MDP). First, we define the state space, which contains key features of the system:

\subsection{System Model}

The state space of the system is defined as follows:
\begin{equation}
\mathbf{s} = \{\mathbf{l}, \mathbf{r}, \mathbf{q}, \mathbf{h}, \mathbf{p}\},
\end{equation}
where $\mathbf{l} \in \mathbb{R}^{d_l}$ represents the current load vector of each service, reflecting real-time request volume and processing capacity; $\mathbf{r} \in \mathbb{R}^{d_r}$ is the resource utilization matrix, containing CPU, memory, and network dimensions; $\mathbf{q} \in \mathbb{R}^{d_g}$ describes the service request queue length; $\mathbf{h} \in \mathbb{R}^{d_h}$ contains historical load statistical features, such as mean and variance; and $\mathbf{p} \in \mathbb{R}^{d_p}$ records system performance metrics, including latency and throughput.

The action space is designed as a composite vector containing multiple scheduling dimensions:
\begin{equation}
\mathbf{a} = \{\mathbf{n}, \mathbf{m}, \mathbf{p}, \mathbf{q}\},
\end{equation}
where $\mathbf{n} \in \mathbb{Z}^k$ represents the adjustment decision for the number of service instances; $\mathbf{m} \in \{0,1\}^{k \times n}$ is the service migration decision matrix, indicating service migration between nodes; $\mathbf{p} \in [0,1]^k$ controls service processing priority; and $\mathbf{q} \in [0,1]^k$ adjusts resource quotas. This multi-dimensional action design enables the scheduler to flexibly respond to different scenarios.

The reward function is designed to comprehensively consider multiple optimization objectives:
\begin{equation}
r_t = - \left( w_1 \sum \frac{T(s_i)}{T_{\text{target}}} + w_2 \sum |u_j - u_{\text{target}}| + w_3 C_t \right),
\end{equation}
The first term penalizes the degree to which service response time exceeds the target threshold, the second term encourages resource utilization to approach the target value, and the third term considers the system overhead caused by scheduling operations. Weight coefficients $w_1$, $w_2$, and $w_3$ are used to balance the importance of each objective.

The optimization objective is to maximize the expected cumulative reward over time:
\begin{equation}
\max_{\pi} \mathbb{E} \left[ \sum_{t=0}^{\infty} \gamma^t R(s_t, a_t) \right],
\end{equation}
where $\pi$ is the policy and $\gamma$ is the discount factor. This objective ensures that the DRL-based scheduler learns a policy that not only optimizes immediate rewards but also considers long-term benefits. The discount factor $\gamma$ balances the trade-off between short-term and long-term rewards, encouraging the model to make decisions that contribute to sustained performance improvements rather than short-lived gains.

The state transition function describes the system dynamics:
\begin{equation}
\mathbf{s}_{t+1} = f(\mathbf{s}_t, \mathbf{a}_t) + \epsilon_t,
\end{equation}
where $f(\cdot)$ simulates the impact of scheduling actions on the system state, and $\epsilon_t$ represents environmental randomness and uncertainty.

To improve policy learning effectiveness, we adopt a Dueling Deep Q-Network (DQN) architecture to design the value function:
\begin{equation}
Q(\mathbf{s}, \mathbf{a}) = V(\mathbf{s}) + A(\mathbf{s}, \mathbf{a}).
\end{equation}


This architecture decomposes the Q-value into a state value function $V(s)$ and an advantage function $A(s, a)$, which can better evaluate the relative advantages of different actions.

Policy optimization uses the Proximal policy optimization (PPO) algorithm \cite{gu2021proximal}, whose objective function is:
\begin{align*}
L = \mathbb{E}_t \left[ \min \left( \frac{\pi_{\theta}(\mathbf{a}_t|\mathbf{s}_t)}{\pi_{\theta_{\text{old}}}(\mathbf{a}_t|\mathbf{s}_t)} \cdot A^{\pi_{\text{old}}}(\mathbf{s}_t, \mathbf{a}_t), \right. \right.
\end{align*}
\begin{equation}
\left. \left. \text{clip} \left( \frac{\pi_{\theta}(\mathbf{a}_t|\mathbf{s}_t)}{\pi_{\theta_{\text{old}}}(\mathbf{a}_t|\mathbf{s}_t)}, 1-\epsilon, 1+\epsilon \right) \cdot A^{\pi_{\text{old}}}(\mathbf{s}_t, \mathbf{a}_t) \right) \right],
\end{equation}
where $\pi_{\theta}$ is the current policy, $\pi_{\theta_{\text{old}}}$ is the old policy, and $\epsilon$ is a small positive constant.

The DRL scheduler is designed to accurately capture the temporal characteristics and bursty patterns of trading loads, automatically balance multiple optimization objectives, adapt to dynamic system changes through continuous learning, and achieve millisecond-level fast scheduling decisions. 

\subsection{Scheduling Algorithm Design}

To achieve efficient resource scheduling, we propose a hybrid scheduling algorithm based on genetic algorithms and reinforcement learning as shown in Alg. \ref{alg:hybrid_scheduling}. The algorithm models service scheduling, resource allocation, and load balancing as a multi-objective optimization problem. The objective function includes three key indicators: service response time $T$, resource utilization $U$, and load balancing degree $L$. The optimization objective function is defined as:
\begin{equation}
f(x) = w_1 \frac{T}{T_{\max}} + w_2 \left(1 - \frac{U}{U_{\max}}\right) + w_3 \left(1 - \frac{L}{L_{\max}}\right),
\end{equation}
where the weight coefficients $w_1 = 0.4$, $w_2 = 0.35$, and $w_3 = 0.25$, and can be customized.

The genetic algorithm iteratively optimizes to obtain the global optimal solution. The algorithm adopts adaptive crossover rate $P_c$ and mutation rate $P_m$:
\begin{equation}
P_c = 0.9 - 0.6 \times \frac{f' - f_{\text{avg}}}{f_{\text{max}} - f_{\text{avg}}},
\end{equation}
\begin{equation}
P_m = 0.1 - 0.07 \times \frac{f' - f_{\text{avg}}}{f_{\text{max}} - f_{\text{avg}}},
\end{equation}
where $f'$ is the current fitness value, $f_{\text{avg}}$ is the average fitness value, and $f_{\text{max}}$ is the maximum fitness value.

In terms of resource allocation strategy optimization, we design a reward function based on reinforcement learning:
\begin{equation}
R(s,a) = \alpha \cdot P(s,a) + \beta \cdot E(s,a) - \gamma \cdot C(s,a),
\end{equation}
where \( \alpha \), \( \beta \), and \( \gamma \) are the weight coefficients that balance the importance of performance gain \( P(s,a) \), resource efficiency \( E(s,a) \), and adjustment cost \( C(s,a) \) in the reward function. The optimization goal is to maximize the cumulative reward \( R \) over time, which reflects the overall effectiveness of the resource allocation strategy.

The algorithm's execution process first initializes a candidate solution population and then enters the iterative optimization phase (lines 6-10). In each iteration, the algorithm generates new scheduling schemes through genetic operations and evaluates and optimizes resource allocation strategies using the reinforcement learning module. The elite solutions are preserved and used to guide the next generation's evolution. The algorithm continues to iterate until convergence conditions or maximum iterations are reached (lines 11-17).

To further improve algorithm efficiency, we adopt several key optimization techniques. First, we use non-dominated sorting to quickly identify high-quality solutions and accelerate population screening. Second, we introduce a local search mechanism to perform fine-tuning near elite solutions (lines 18-23). Finally, we adopt adaptive population size control to dynamically adjust computational resource allocation during algorithm execution (lines 24-26).


The main advantages of this hybrid optimization method are: it can simultaneously consider global search and local optimization, has strong environmental adaptability, and can find a good balance between multiple optimization objectives. The algorithm's adaptive characteristics enable it to effectively respond to dynamic changes in system load, providing a feasible solution for resource scheduling in microservice architectures.

\begin{algorithm}
	\caption{Hybrid Resource Scheduling Algorithm}
	\label{alg:hybrid_scheduling}
	\SetKwProg{Fn}{Function}{:}{end}
	\SetAlgoLined
	\DontPrintSemicolon 
	\LinesNumbered
	\SetAlgoVlined
	
	\Fn{\textsc{HybridScheduling}(S, R, T)}{
		$P \leftarrow$ \textsc{GenerateInitialPopulation}(N)\;
		$state \leftarrow$ \textsc{GetSystemState}()\;
		$best\_solution \leftarrow$ null\;
		
		\While{iteration $<$ MAX\_ITER \textbf{and} not Converged}{
			\For{each $x \in P$}{
				fitness[$x$] $\leftarrow$ $w_1 \left(\frac{T}{T_{\max}}\right) + w_2 \left(1 - \frac{U}{U_{\max}}\right) + w_3 \left(1 - \frac{L}{L_{\max}}\right)$\;
			}
			$P_c \leftarrow 0.9 - 0.6 \times \frac{f' - f_{\text{avg}}}{f_{\max} - f_{\text{avg}}}$\;
			$P_m \leftarrow 0.1 - 0.07 \times \frac{f' - f_{\text{avg}}}{f_{\max} - f_{\text{avg}}}$\;
			elite $\leftarrow$ \textsc{SelectTopK}($P$, fitness, $k$)\;
			\For{solution in elite}{
				action $\leftarrow$ \textsc{PolicyNetwork}(state)\;
				reward $\leftarrow \alpha \cdot P(s,a) + \beta \cdot E(s,a) - \gamma \cdot C(s,a)$\;
				next\_state $\leftarrow$ \textsc{ExecuteAction}(action)\;
				\textsc{UpdatePolicyNetwork}(state, action, reward, next\_state)\;
				state $\leftarrow$ next\_state\;
			}
			$offspring$ $\leftarrow \emptyset$\;
			\While{$|\text{offspring}| < N-k$}{
				parents $\leftarrow$ \textsc{TournamentSelect}($P$, 2)\;
				$child_1$, $child_2$ $\leftarrow$ \textsc{Crossover}(parents, $P_c$)\;
				$child_1$ $\leftarrow$ \textsc{Mutation}($child_1$, $P_m$)\;
				$child_2$ $\leftarrow$ \textsc{Mutation}($child_2$, $P_m$)\;
				$offspring$ $\leftarrow$ $offspring$ $\cup$ \{$child_1$, $child_2$\}\;
			}
			$P \leftarrow$ $elite$ $\cup$ $offspring$\;
			\If{fitness[best($P$)] $>$ fitness[$best\_solution$]}{
				$best\_solution \leftarrow$ best($P$)\;
			}
		}
		\Return $best\_solution$\;
	}
\end{algorithm}

\subsection{Improvement Scheme for Adaptive Trading System}
The algorithm also considers improvement process as follows:

\subsubsection{Adaptive Resource Scheduling Algorithm} Aiming at the bursty traffic characteristics of trading systems, an adaptive resource scheduling algorithm based on deep
reinforcement learning was designed. The algorithm represents the system state as a vector $S=(C,M,N,L)$, where $C$ is the
CPU utilization, $M$ is the memory occupancy, $N$ is the network throughput, and $L$ is the load level. The action space $A$
includes container scaling, resource quota adjustment, and other operations. Through deep Q-network training, the
optimal scheduling strategy is achieved, realizing an adaptive resource allocation mechanism. The algorithm monitors
the system state in real-time and makes rapid scheduling decisions based on the pre-set strategy model, ensuring a
balance between resource utilization efficiency and system performance.

\subsubsection{Multi-Level Cache Mechanism} A three-level cache architecture was constructed for trading systems, as shown in Fig. \ref{fig:sys_arch}, including local memory cache
(L1), distributed cache (L2), and persistent storage (L3). L1 cache uses the least recently used (LRU) algorithm \cite{zhang2025incremental} to manage active trading
data, L2 cache uses Redis clusters to store hot data, and achieves load balancing through consistent hashing, while L3
uses distributed storage to ensure data persistence. The specific configuration parameters for each cache level
are shown in Table \ref{tab:cache}. The multi-version concurrency control (MVCC) protocol ensures cache consistency, and differentiated expiration strategies are set for
different cache levels. This mechanism effectively reduces system response time and improves memory usage efficiency.

\begin{table}[ht!]
\caption{Cache Level Specifications}
\label{tab:cache}
\centering
\begin{tabular}{|c|c|c|c|}
\hline
\textbf{Level} & \textbf{Algorithm} & \textbf{Expiration Time} & \textbf{Consistency} \\
\hline
L1 & LRU & 10s & MVCC \\
L2 & Consistent Hash & 60s & MVCC \\
L3 & -- & Permanent & Strong \\
\hline
\end{tabular}
\end{table}

\subsubsection{Load Prediction Model} A load prediction model based on LSTM deep learning networks was constructed for the trading system, with its architecture illustrated in Fig. \ref{fig:sys_arch}. The model utilizes 18 input features, including historical trading volume, time-specific features, and market indicators. The model comprises 3 layers of LSTM structures, each with 128 neurons and a dropout rate of 0.3. Training is conducted using the Adam optimizer, with mean squared error (MSE) as the loss function. This model effectively predicts future trading volumes and provides early warnings for bursty traffic, offering a decision-making foundation for system resource scheduling and significantly enhancing the stability of trading systems.

\section{Experimental Evaluations}
In this section, we introduce our experimental settings as well as the observed results.

\subsection{Experimental Setup}

\subsubsection{Test Environment Configurations}The experiments were conducted in a standard cloud computing test environment. The hardware platform consisted of 20 high-performance servers, each equipped with an Intel Xeon Gold 6248R processor (3.0 GHz, 24 cores, 48 threads), 384 GB DDR4 memory, and a 4TB NVMe solid-state drive. The network environment featured a 10 Gbps Ethernet interconnection with latency under 0.5ms. As detailed in Table \ref{tab:config}, the operating system used was Ubuntu 20.04 LTS, with kernel version 5.15.0, Docker version 20.10.12, and Kubernetes version 1.24.3. Monitoring tools such as Prometheus 2.36.0 and Grafana 9.0.2 were employed, with a sampling interval of 10 seconds and a data retention period of 30 days.

\begin{table}[ht!]  
\centering
\caption{Configuration Items and Parameter Specifications}
\label{tab:config}
\begin{tabular}{|l|l|}
\toprule
\textbf{Configuration Item} & \textbf{Parameter Specification} \\
\midrule
CPU & Intel Xeon Gold 6248R, 3.0GHz, 24C48T \\
Memory & 384GB DDR4-2933 \\
Storage & 4TB NVMe SSD \\
Network & 10Gbps Ethernet \\
OS & Ubuntu 20.04 LTS \\
Container Runtime & Docker 20.10.12 \\
Orchestration & Kubernetes 1.24.3 \\
\bottomrule
\end{tabular}
\end{table}

\subsubsection{Online Trading System Test Scenario}
\begin{figure}[ht!]
    \centering
    \includegraphics[width=0.95\linewidth]{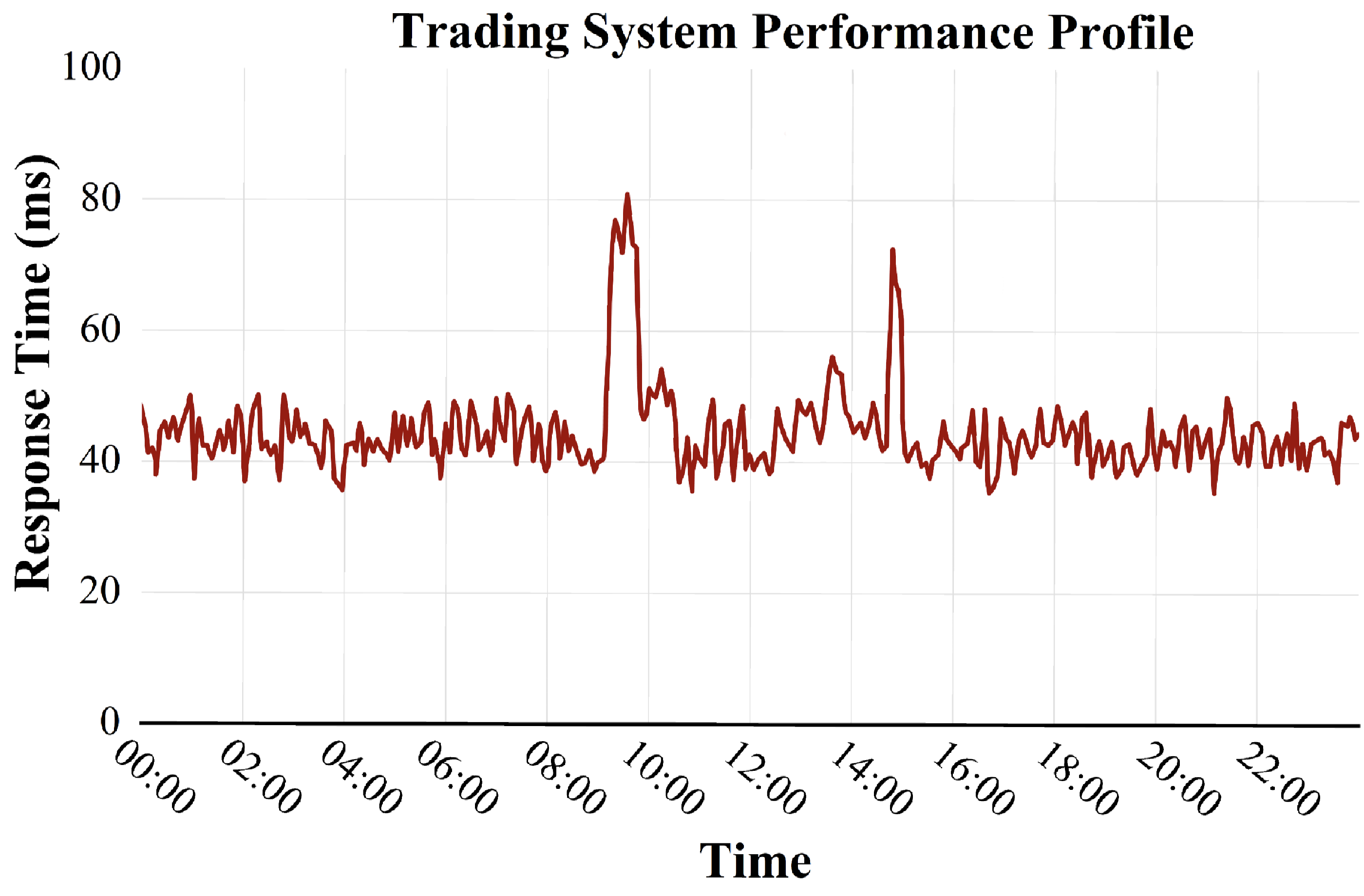}
    \caption{Trading System Running Status (Response Time).}
    \label{fig:Trading_System1}
\end{figure}

\begin{figure}[ht!]
    \centering
    \includegraphics[width=0.95\linewidth]{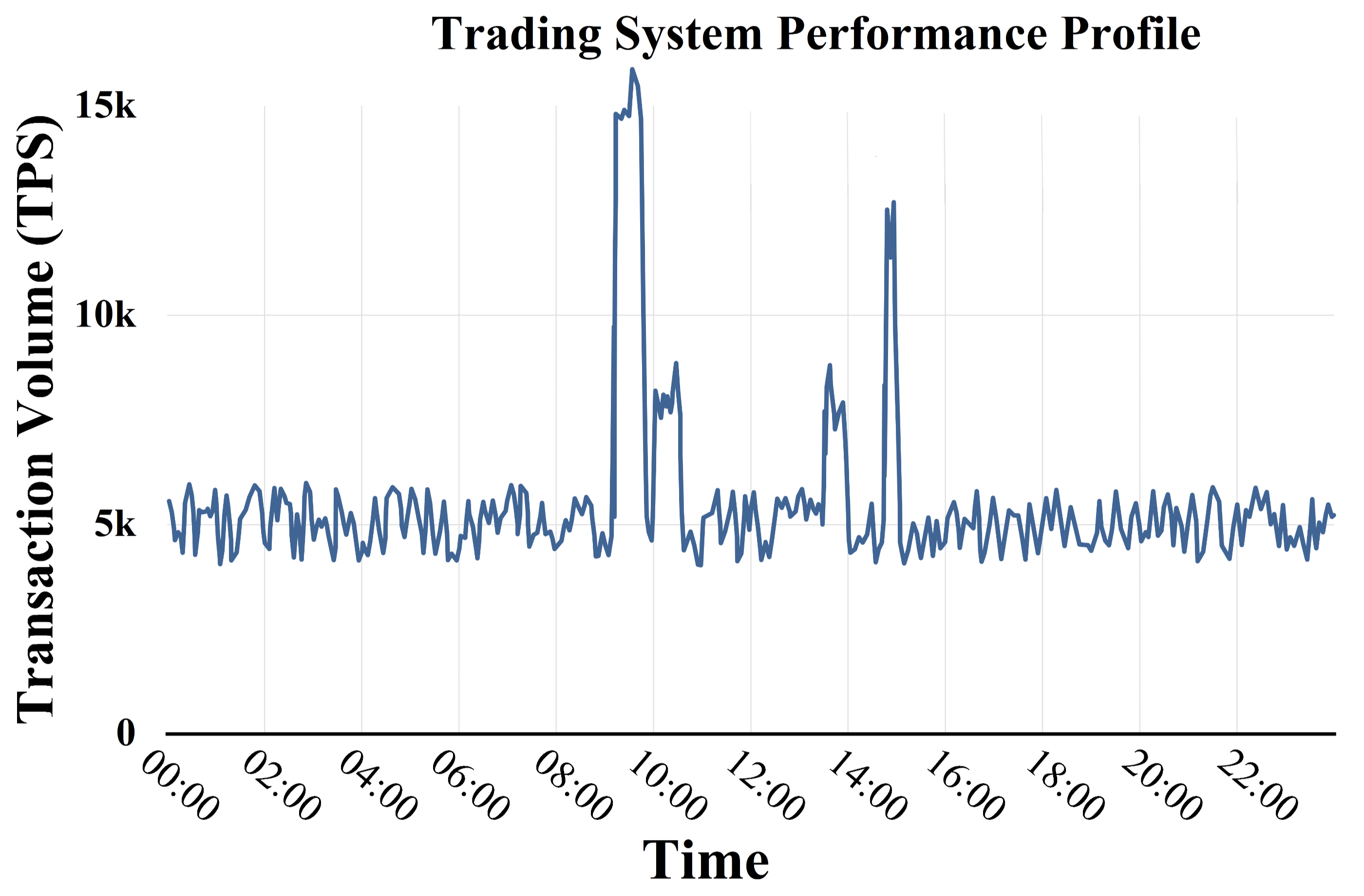}
    \caption{Trading System Running Status (Transaction Volume).}
    \label{fig:Trading_System2}
\end{figure}

A production-level trading system from a certain securities company was selected as the benchmark test application. The system consisted of 8 core microservice modules, including account authentication, market data push, trading commission, order matching, clearing and settlement, risk control monitoring, ledger management, and message notification. Each service module was deployed with 3-5 instances to ensure high availability, totaling 32 service instances. The system's performance, including response time and transaction volume, are shown in Fig. \ref{fig:Trading_System1} and Fig. \ref{fig:Trading_System2}.

A JMeter pressure test model was employed to simulate the market opening and closing periods. The test load followed the characteristics of financial transactions, where the number of concurrent users increased from 1,000 to 10,000 over 15 minutes before the market opened. Transaction requests adhered to a Poisson distribution, reflecting the bursty nature of trading volume. The test dataset included 5 million active trading accounts, 2,000 stock targets, and 80 million historical order records, simulating a real trading environment with a total data size of 2TB. The system was required to handle 5,000 transactions per second (TPS) under normal conditions and 15,000 TPS during peak periods, with an average processing time of no more than 50ms per transaction to meet the exchange's performance requirements. The risk control rules involved more than 120 real-time monitoring indicators, such as trading limits and price limits, and each transaction request underwent verification through the entire risk control chain.

\subsection{Performance Comparison}
\subsubsection{Comparison with Virtual Machine (VM) Deployment Scheme} 
\begin{figure*}[htbp]
    \centering
    \begin{subfigure}[b]{0.32\textwidth}
        \centering
        \includegraphics[width=\textwidth]{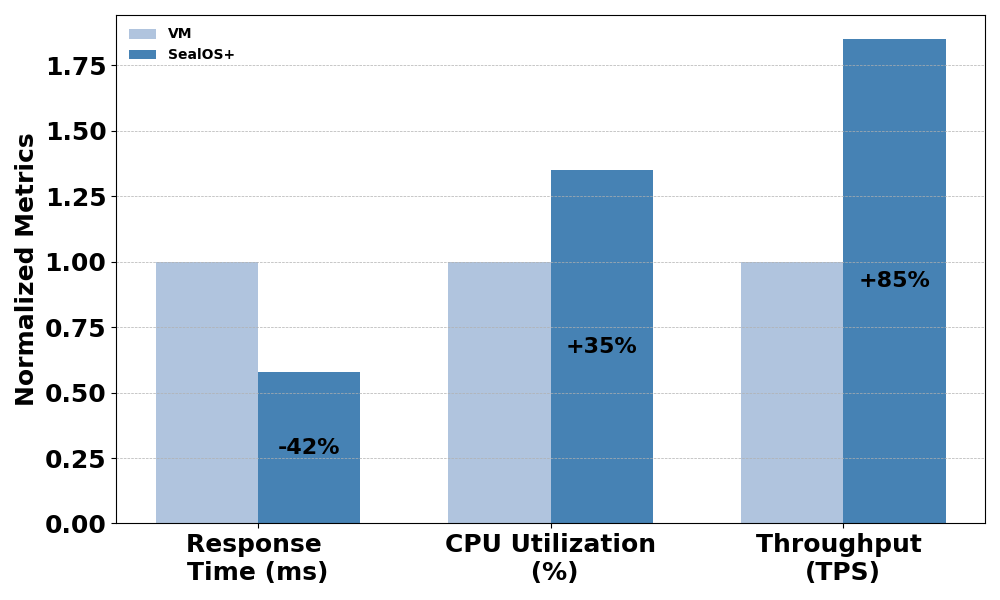}
        \caption{Performance Comparison Between SealOS+ and VM Solution.}
        \label{fig:con vs vm}
    \end{subfigure}
    \hfill
    \begin{subfigure}[b]{0.32\textwidth}
        \centering
        \includegraphics[width=\textwidth]{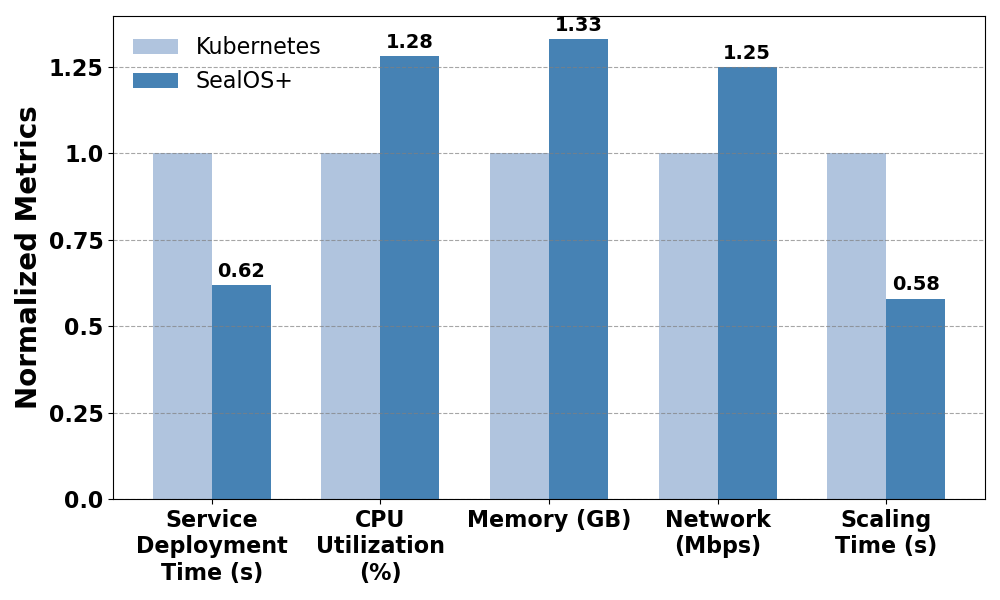}
        \caption{Performance Comparison Between SealOS+ and Kubernetes.}
        \label{fig:Performance_k8s}
    \end{subfigure}
    \hfill
    \begin{subfigure}[b]{0.32\textwidth}
        \centering
        \includegraphics[width=\textwidth]{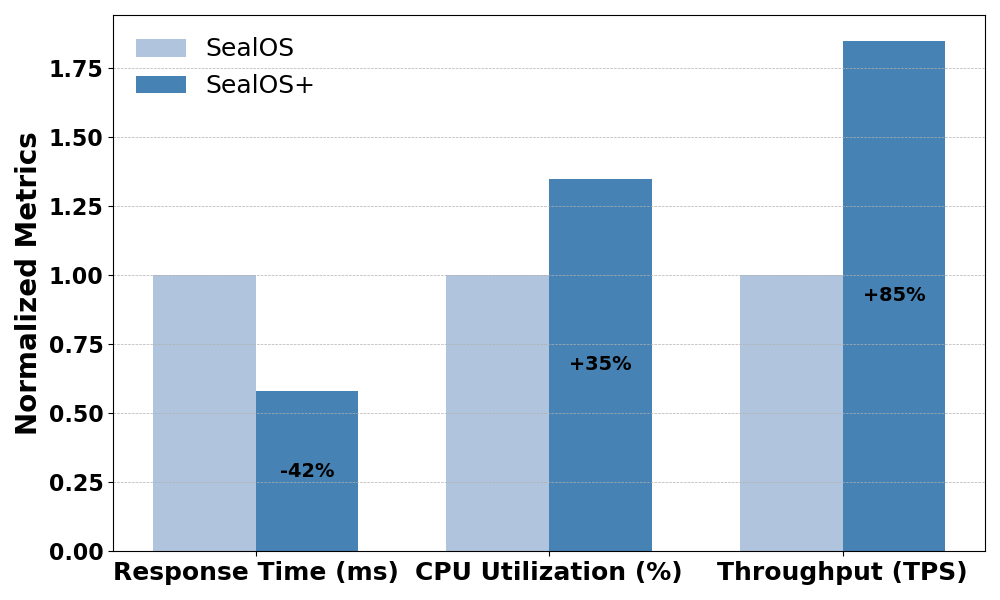}
        \caption{Performance Comparison Between SealOS+ and SealOS.}
        \label{fig:Performance_sealos}
    \end{subfigure}
    \caption{SealOS+ performance compared with VM solution, Kubernetes and SealOS}
    \label{fig:sealOS+}
\end{figure*}

To show the benefits of containerized solution, a comparison analysis was conducted between SealOS+ and the traditional VM deployment scheme for the same microservice application scenario. The performance differences were evaluated across three key indicators: CPU utilization, memory usage efficiency, and disk I/O performance, as shown in Fig. \ref{fig:con vs vm}. The containerized SealOS+ scheme achieved an average CPU utilization of 72\%, which is 23\% higher than the VM scheme. Memory usage efficiency improved by 35\%, with a single node now able to deploy 28 service instances, compared to 12 instances in the VM setup. Disk I/O performance also saw a 52\% improvement, with the average response time reduced to 65ms, down from 150ms in the VM scheme. Furthermore, system startup time was significantly reduced from 180 seconds in the VM setup to just 25 seconds, resulting in an 86\% improvement in resource scheduling efficiency.

\subsubsection{Comparison with Native Kubernetes} To evaluate the performance compared to native Kubernetes, a comparison analysis was conducted between the SealOS+ and native Kubernetes in a large-scale cluster scenario, focusing on service scheduling efficiency, resource utilization, and system scalability, as shown in Fig. \ref{fig:Performance_k8s}. Test data revealed that the improved scheme optimized service deployment time from 85 seconds in native Kubernetes to 32 seconds, marking a 62\% improvement. The resource allocation algorithm boosted cluster-wide CPU utilization by 28\%, reaching 78\%. Additionally, memory usage efficiency increased by 33\%, and network bandwidth utilization improved by 25\%. The system can now stably support 5,000 service instances running simultaneously, with a 58\% reduction in scaling operation time.

\subsubsection{Comparison with default Sealos} To demonstrate the performance improvement with default SealOS,  a  performance comparison was conducted between the SealOS+ and the SealOS as shown in Table \ref{tab:performance_sealos}, with pressure testing focused on core functions such as service discovery, load balancing, and fault recovery. The results showed that the optimized service discovery mechanism reduced registration time from 45ms to 19ms, and query latency from 65ms to 29ms. The new load balancing algorithm minimized request allocation imbalance from 23\% to 8\%, while fault recovery time was shortened from 90 seconds to 35 seconds. As shown in Table \ref{tab:performance_sealos}, in high-concurrency scenarios, system throughput increased by 85\%, reaching 25,000 QPS, while response time was reduced by 62\%, and service availability improved from 99.95\% to 99.99\%.

\begin{table}[ht!]
\centering
\caption{Sealos Optimization Effect Comparison.}
\label{tab:performance_sealos}
\begin{tblr}{hlines, vlines,
  colspec={*{4}{X[c]}},
  rowsep=0.05cm, colsep=0.1cm
}
\textbf{Metrics} & \textbf{SealOS} & \textbf{SealOS+} & \textbf{Improvement}\\
Service Reg. Time & 45ms & 19ms & \textbf{60\%} \\
Load Imbalance & 23\% & 8\% & \textbf{65\%} \\
Fault Recovery Time & 90s & 35s & \textbf{60\% } \\
System Throughput & 13,500 TPS & 25,000 TPS & \textbf{85\%} \\
Service Availability & 99.95\% & 99.99\% & \textbf{0.04\%}
\end{tblr}
\end{table}

Through deployment testing in a securities trading system production environment, the system's core indicators showed significant improvements compared to the original SealOS version, as illustrated in Fig. \ref{fig:Performance_sealos}. During the opening collective bidding phase, the system's average response time was reduced from 180ms to 105ms, marking a 42\% decrease, effectively meeting the exchange's stringent requirements for trading latency. The improved resource scheduling algorithm enhanced node CPU utilization from 45\% to 78\%, a 35\% increase, significantly boosting hardware resource utilization efficiency while ensuring system stability. Most notably, the system's peak processing capacity rose from 8,000 TPS to 14,800 TPS, an 85\% increase, providing ample performance reserves to handle market fluctuations.


\subsection{Latency Analysis}

\subsubsection{End-to-End Service Response Time}
\begin{figure}[ht!]
    \centering
    \includegraphics[width=0.97\linewidth]{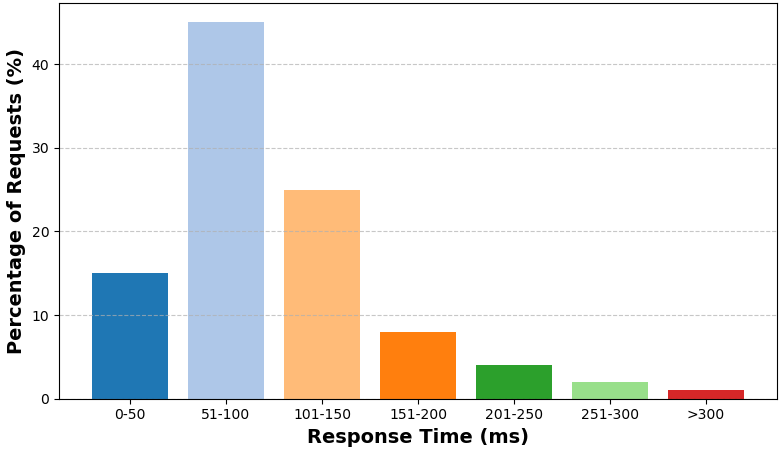}
    \caption{End-to-End Response Time.}
    \label{fig:End-to-End}
\end{figure}

As shown in Fig. \ref{fig:End-to-End}, statistical analysis of 10 million request samples revealed that the system's end-to-end service response latency followed a lognormal distribution with an average of 85ms under normal load (5000 TPS). Specifically, 95\% of requests responded within 120ms and 99\% within 150ms, with a standard deviation of 29ms. The latency breakdown included network transmission (15ms), service processing (45ms), and database access (25ms). Using Jaeger for distributed tracing, it was found that gateway processing and service discovery accounted for 18\% of the total time, highlighting areas for optimization.

\subsubsection{Latency Performance under Different Loads}
A stepped load test evaluated the system's latency performance across varying concurrency levels. As shown in Table \ref{SystemLatency}, the average response latency increased from 95ms at 2000 concurrent users to 285 
ms at 10000 concurrent users. The system maintained stable performance below 6000 concurrent users, but latency rose significantly above this threshold. Monitoring revealed that the database connection pool became a bottleneck under high load. Optimization efforts reduced the maximum latency by 32\%.

\begin{table}[htbp]
\caption{System Latency Performance under Different Concurrency Levels.}
\label{SystemLatency}
\centering
\begin{tblr}{hlines, vlines,
  colspec={*{6}{X[c]}},
  rowsep=0.05cm, colsep=0.05cm
}
\textbf{Users} & \textbf{Avg RT (ms)} & \textbf{95\% RT (ms)} & \textbf{99\% RT (ms)} & \textbf{CPU Util. (\%)} & \textbf{Mem Util. (\%)}\\
2000 & 95 & 135 & 165 & 45 & 52\\
4000 & 125 & 178 & 225 & 65 & 68\\
6000 & 168 & 245 & 298 & 78 & 75\\
8000 & 215 & 312 & 385 & 86 & 82\\
10000 & 285 & 425 & 520 & 92 & 88\\
\end{tblr}
\end{table}

\subsection{ System Overhead Analysis}

\subsubsection{Resource Utilization} 
\begin{figure}[ht!]
    \centering
    \includegraphics[width=0.94\linewidth]{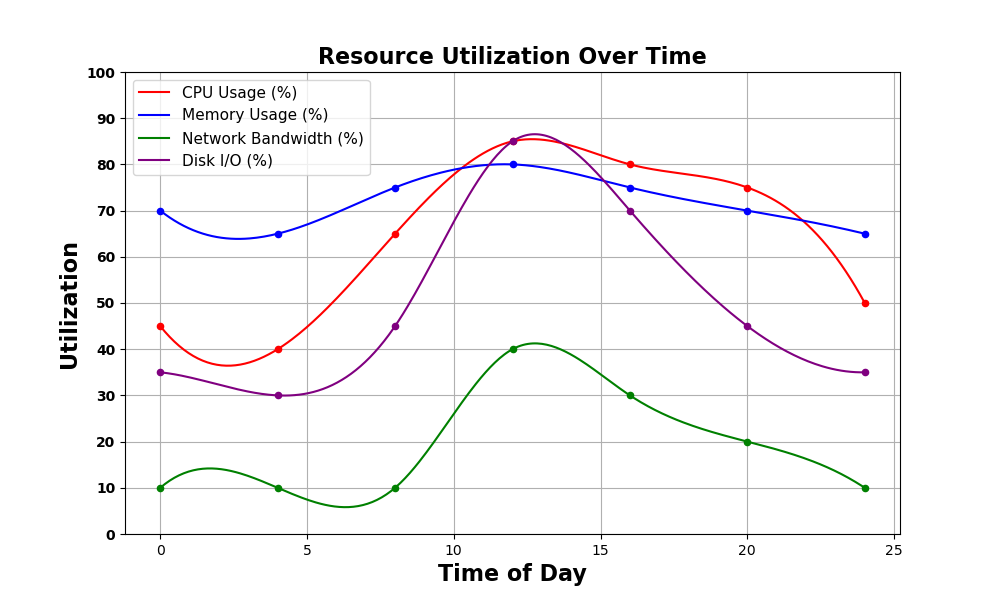}
    \caption{Resource Utilization Over Time.}
    \label{fig:Resource Utilization Over Time}
\end{figure}

Through the analysis of monitoring data from a 20-node cluster in a production environment over a 30-day period, detailed resource usage metrics were recorded. As shown in Fig. \ref{fig:Resource Utilization Over Time}, under standard workload conditions, the CPU average utilization remained around 65\%, with a peak of 88\% during peak hours. Memory usage was relatively stable, averaging 72\%, with fluctuations between 65\% and 78\%. Network bandwidth utilization exhibited a clear tidal pattern, averaging 4.2 Gbps during the day and 1.8 Gbps at night. Disk I/O usage stayed below 45\% during normal operations, with short-term peaks reaching 85\% during data backups. Resource utilization efficiency improved by 35\% compared to SealOS, with the number of service instances supported per unit of energy consumption increasing from 85 to 115. Each service instance averaged 2.3 CPU cores and 3.8 GB of memory, demonstrating efficient resource usage.

\subsubsection{System Load} An in-depth analysis was performed on the system load, collecting key indicators such as average load, process count, and context switch count. As shown in Table \ref{tab:System Load}, under normal operating conditions, the system's 1-minute average load was 4.2, the 5-minute average load was 3.8, and the 15-minute average load was 3.5, indicating effective load balancing. The context switch count per second remained around 25,000, representing a 32\% decrease compared to the native Kubernetes cluster. With the optimized scheduling algorithms, the process count was maintained within 780 per node, and the interrupt handling time ratio decreased from 12\% to 7.5\%. In the event of sudden traffic surges, the system's load growth rate was controlled within 35\%, with service response time increasing by no more than 25\%, demonstrating excellent load-bearing capacity.

\begin{table}[ht!]
\caption{System Load Key Indicator Statistics}
\label{tab:System Load}
\centering
\begin{tblr}{hlines, vlines,
  colspec={*{6}{X[c]}},
  rowsep=0.05cm, colsep=0.1cm
}
\textbf{Time Period} & \textbf{Average Load} & \textbf{Process Count} & \textbf{Context Switch} & \textbf{CPU Utilization} \\
Morning Peak & 5.8 & 750 & 28.5k & 78\% \\
Normal Period & 4.2 & 620 & 25k & 65\% \\
Nighttime & 2.5 & 480 & 18.5k & 45\%  \\
Sudden Traffic & 7.2 & 780 & 32k & 85\% \\
Daily Average & 4.2 & 650 & 25k & 68\% 
\end{tblr}
\end{table}

To summarize, SealOS+ can achieve significant performance improvement compared with VM solution, Kubernetes and default SealOS in terms of resource utilization and response time. 

\section{Conclusions}

This paper presents a comprehensive performance optimization solution for securities trading systems using SealOS+, an enhanced container orchestration platform. The proposed system addresses key challenges in trading environments, such as network communication overhead, resource scheduling inefficiencies, and storage latency. By integrating an adaptive batch scheduling algorithm and a load prediction model, the solution significantly improves system performance and reliability. The adaptive scheduling reduces trading chain processing time from 150ms to 40ms, meeting the 50ms requirement while supporting peak transaction volumes of 15,000 TPS. The load prediction model forecasts trading volume changes with 92\% accuracy, while optimized resource scheduling saves 30\% in computing resources. Additionally, a new batch caching mechanism reduces database access frequency by 90\%, speeding up account processing for clearing and risk control. 

In conclusion, the proposed solution optimizes resource utilization, minimizes latency, and improves system scalability, making it well-suited for high-frequency trading scenarios. Future work will focus on extending the solution to futures and fund trading systems and validating it on larger clusters to support even higher transaction volumes. The optimization on network costs will also be considered.  

\section*{Acknowledgments}
This work is supported by Guangdong Basic and Applied Basic Research Foundation (No. 2024A1515010251, 2023B1515130002), Guangdong Special Support Plan (No. 2021TQ06X990), Shenzhen Basic Research Program under grants JCYJ20220818101610023, and JCYJ20240809180935001, Shenzhen Industrial Application Projects of undertaking the National key R \& D Program of China (No. CJGJZD20210408091600002).

\bibliographystyle{unsrt}
\bibliography{sample}

\begin{thebibliography}{10}

\bibitem{zhong2024domain}
Chenxing Zhong, Shanshan Li, Huang Huang, Xiaodong Liu, Zhikun Chen, Yi~Zhang,
  and He~Zhang.
\newblock Domain-driven design for microservices: An evidence-based
  investigation.
\newblock {\em IEEE Transactions on Software Engineering}, 2024.

\bibitem{bai2024DRPC}
Haoyu Bai, Minxian Xu, Kejiang Ye, Rajkumar Buyya, and Chengzhong Xu.
\newblock Drpc: Distributed reinforcement learning approach for scalable
  resource provisioning in container-based clusters.
\newblock {\em IEEE Transactions on Services Computing}, 2024.

\bibitem{sealos}
Sealos.
\newblock {Sealos: Develop, deploy, and scale in one seamless cloud platform}.
\newblock \url{https://sealos.io/}, 2024.
\newblock [Online; accessed 13-March-2025].

\bibitem{cite2}
Guangsheng Yu, Xu~Wang, Wei Ni, and et~al.
\newblock Adaptive resource scheduling in permissionless sharded-blockchains: A
  decentralized multiagent deep reinforcement learning approach.
\newblock {\em IEEE Transactions on Systems, Man, and Cybernetics: Systems},
  53(11):7256--7268, 2023.

\bibitem{cite3}
Thomas Fischer and Christopher Krauss.
\newblock Deep learning with long short-term memory networks for financial
  market predictions.
\newblock {\em European Journal of Operational Research}, 270(2):654--669,
  2018.

\bibitem{yu2019review}
Yong Yu, Xiaosheng Si, Changhua Hu, and Jianxun Zhang.
\newblock A review of recurrent neural networks: Lstm cells and network
  architectures.
\newblock {\em Neural computation}, 31(7):1235--1270, 2019.

\bibitem{cite4}
Senthil~G. A, K.~Somasundaram, Arun M, N.~Naga Saranya, R.~Prabha, and
  D.~Vijendra Babu.
\newblock A novel hybrid gaaco algorithm for cloud computing using energy aware
  load balance scheduling.
\newblock In {\em 2022 International Conference on Computer Communication and
  Informatics (ICCCI)}, pages 1--5, 2022.

\bibitem{cite12}
Nguyen Nguyen and Taehong Kim.
\newblock Toward highly scalable load balancing in kubernetes clusters.
\newblock {\em IEEE Communications Magazine}, 58(7):78--83, 2020.

\bibitem{cite16}
Byeonghui Jeong and Young-Sik Jeong.
\newblock Arascaler: Adaptive resource autoscaling scheme using etimemixer for
  efficient cloud-native computing.
\newblock {\em IEEE Transactions on Services Computing}, PP:1--14, 01 2024.

\bibitem{cite5}
Yeddula Sai~Dhanush Reddy, Padumati~Saikiran Reddy, Nithya Ganesan, and
  B.~Thangaraju.
\newblock Performance study of kubernetes cluster deployed on openstack,vms and
  baremetal.
\newblock In {\em 2022 IEEE International Conference on Electronics, Computing
  and Communication Technologies (CONECCT)}, pages 1--5, 2022.

\bibitem{cite6}
Angelo Marchese and Orazio Tomarchio.
\newblock {\em Network SLO-Aware Container Orchestration on Kubernetes
  Clusters}, pages 96--104.
\newblock 12 2024.

\bibitem{cite14}
Carmen Carri\'{o}n.
\newblock Kubernetes scheduling: Taxonomy, ongoing issues and challenges.
\newblock {\em ACM Comput. Surv.}, 55(7), December 2022.

\bibitem{cite13}
Emiliano Casalicchio.
\newblock {\em Container Orchestration: A Survey}, pages 221--235.
\newblock Springer International Publishing, Cham, 2019.

\bibitem{cite8}
Minxian Xu and Rajkumar Buyya.
\newblock Brownout approach for adaptive management of resources and
  applications in cloud computing systems: A taxonomy and future directions.
\newblock {\em ACM Comput. Surv.}, 52(1), January 2019.

\bibitem{cite10}
Zhiheng Zhong, Minxian Xu, Maria~Alejandra Rodriguez, Chengzhong Xu, and
  Rajkumar Buyya.
\newblock Machine learning-based orchestration of containers: A taxonomy and
  future directions.
\newblock {\em ACM Comput. Surv.}, 54(10s), September 2022.

\bibitem{cite11}
Zhaolong Jian, Xueshuo Xie, Yaozheng Fang, Yibing Jiang, Ye~Lu, Ankan Dash, Tao
  Li, and Guiling Wang.
\newblock Drs: A deep reinforcement learning enhanced kubernetes scheduler for
  microservice-based system.
\newblock {\em Software: Practice and Experience}, 54(10):2102--2126, 2024.

\bibitem{cite15}
Zhao Li, Haifeng Sun, Zheng Xiong, Qun Huang, Zehong Hu, Ding Li, Shasha Ruan,
  Hai Hong, Jie Gui, Jintao He, Zebin Xu, and Yang Fang.
\newblock Noah: Reinforcement-learning-based rate limiter for microservices in
  large-scale e-commerce services.
\newblock {\em IEEE transactions on neural networks and learning systems}, PP,
  04 2023.

\bibitem{cite17}
Kaiwen Li, Tao Zhang, and Rui Wang.
\newblock Deep reinforcement learning for multiobjective optimization.
\newblock {\em IEEE Transactions on Cybernetics}, 51(6):3103--3114, 2021.

\bibitem{chen2024scaling}
Austin Chen and Genya Ishigaki.
\newblock Scaling container caching to larger networks with multi-agent
  reinforcement learning.
\newblock In {\em 2024 33rd International Conference on Computer Communications
  and Networks (ICCCN)}, pages 1--5. IEEE, 2024.

\bibitem{gu2021proximal}
Yang Gu, Yuhu Cheng, CL~Philip Chen, and Xuesong Wang.
\newblock Proximal policy optimization with policy feedback.
\newblock {\em IEEE Transactions on Systems, Man, and Cybernetics: Systems},
  52(7):4600--4610, 2021.

\bibitem{zhang2025incremental}
Jinbei Zhang, Chunpeng Chen, Kechao Cai, and John~CS Lui.
\newblock Incremental least-recently-used algorithm: Good, robust, and
  predictable performance.
\newblock {\em IEEE Transactions on Mobile Computing}, 2025.

\end{thebibliography}

\end{document}